\documentclass[conference]{IEEEtran}
\usepackage[left=0.625in, right=0.625in, bottom=0.9in, top=0.65in]{geometry}

\makeatletter
\def\endthebibliography{%
	\def\@noitemerr{\@latex@warning{Empty `thebibliography' environment}}%
	\endlist
}
\makeatother

\usepackage{amsfonts}
\usepackage{times}
\usepackage{graphicx}
\usepackage{epstopdf}
\usepackage{latexsym}
\usepackage{dsfont}
\usepackage{amssymb}
\usepackage{amsmath}
\usepackage{cite}
\usepackage{verbatim}

\newcommand{\figref}[1]{{Fig.}~\ref{#1}}

\def\bb0{{\mathbb{0}}}

\def\bb{{\mathbf{b}}}

\def\bff{{\mathbf{f}}}

\def\bh{{\mathbf{h}}}

\def\bp{{\mathbf{p}}}

\def\bu{{\mathbf{u}}}

\def\b0{{\mathbf{0}}}

\def\bH{{\mathbf{H}}}

\def\bbC{{\mathbb{C}}}

\def\bbE{{\mathbb{E}}}

\def\bbR{{\mathbb{R}}}

\def\cC{\mathcal{C}}

\def\cE{\mathcal{E}}
\def\cF{\mathcal{F}}

\def\cH{\mathcal{H}}

\def\cN{\mathcal{N}}

\def\cS{\mathcal{S}}
\def\cT{\mathcal{T}}

\def\sf0{{\mathsf{0}}}

\usepackage{graphicx}
\usepackage{epstopdf}
\usepackage{enumerate}
\usepackage{algorithm}
\usepackage{amsmath}
\usepackage{mathrsfs}
\usepackage{float}
\usepackage{color}
\usepackage{makeidx}
\usepackage{bm}
\usepackage{cleveref}
\usepackage{url}
\usepackage{steinmetz}
\usepackage{varwidth}% http://ctan.org/pkg/varwidth
\usepackage{soul}
\usepackage{bbm}
\usepackage{empheq}
\usepackage{mathtools}
\usepackage{cite}
\usepackage{balance}
\usepackage{multirow}
\usepackage{subfigure}
\usepackage{epstopdf}
\newcommand{\sref}[1]{{Section}~\ref{#1}}

\DeclareMathOperator*{\argmax}{arg\,max}

\def \rm {\mathrm}

\usepackage{amsfonts}
\usepackage{booktabs}
\usepackage{siunitx}

\begin{document}
	\title{\huge Digital Twin Based Beam Prediction:\\ Can we Train in the Digital World and Deploy in Reality?}
	\author{Shuaifeng Jiang and Ahmed Alkhateeb\\ \textit{School of Electrical, Computer and Energy Engineering - Arizona State University} \\ \textit{Emails: \{s.jiang, alkhateeb\}@asu.edu}\thanks{This work is supported by the National Science Foundation under Grant No. 2048021.}}
	
	\maketitle
	\begin{abstract}
		Realizing the potential gains of large-scale MIMO systems requires the accurate estimation of their channels or the fine adjustment of their narrow beams. This, however, is typically associated with high channel acquisition/beam sweeping overhead that scales with the number of antennas. Machine and deep learning represent promising approaches to overcome these challenges thanks to their powerful ability to learn from prior observations and side information. Training machine and deep learning models, however, requires large-scale datasets that are expensive to collect in deployed systems. To address this challenge, we propose a novel direction that utilizes digital replicas of the physical world to reduce or even eliminate the MIMO channel acquisition overhead. In the proposed digital twin aided communication, 3D models that approximate the real-world communication environment are constructed and accurate ray-tracing is utilized to simulate the site-specific channels. These channels can then be used to aid various communication tasks. Further, we propose to use machine learning to approximate the digital replicas and reduce the  ray tracing computational cost. To evaluate the proposed digital twin based approach, we conduct a case study focusing on the position-aided beam prediction task. The results show that a learning model trained solely with the data generated by the digital replica can achieve relatively good performance on the real-world data. Moreover, a small number of real-world data points can quickly achieve near-optimal performance, overcoming the modeling mismatches between the physical and digital worlds and significantly reducing the data acquisition overhead.
	\end{abstract}


	\section{Introduction}\label{Introduction}
	Multiple-input and Multiple-Output (MIMO) communication gains have been widely investigated in the last three decades. With the multiplexing and array gains of these systems, scaling the number of antennas up is considered as a key characteristic of current and future wireless communication systems in 5G and beyond \cite{Boccardi2014}. 
	Realizing these gains, however, requires some knowledge about the channels of which the acquisition overhead scales with the number of antennas \cite{Boccardi2014}. The channel estimation/feedback (and beam sweeping) overhead makes it hard for these systems to continue scaling their gains or to support highly-mobile applications. 
	Machine and deep learning provide a promising path for overcoming these challenges by leveraging the prior observations to reduce the channel/beam acquisition overhead. Training machine/deep learning models, however, requires large-scale datasets that are hard to collect in reality without penalizing the system performance.

	In this paper, \textbf{we propose a novel direction that exploits the digital twin to aid wireless communication systems with reduced channel acquisition overhead}. The proposed digital twin-based communications assume that the BSs have information on the position, orientation, dynamics, shapes, and materials of the surrounding objects. With this information, the BSs then construct 3D models of the surrounding environments. Using the 3D model and ray tracing, the BSs can simulate the channels for each communication link in a digital replica of the real world. This simulated channel information can then be exploited to aid various communication tasks in the real world. If the digital replica is adequately accurate, the real-world channel acquisition can even be bypassed.
	
		\begin{figure*}[t]
		\centering
		\includegraphics[width=0.9\linewidth]{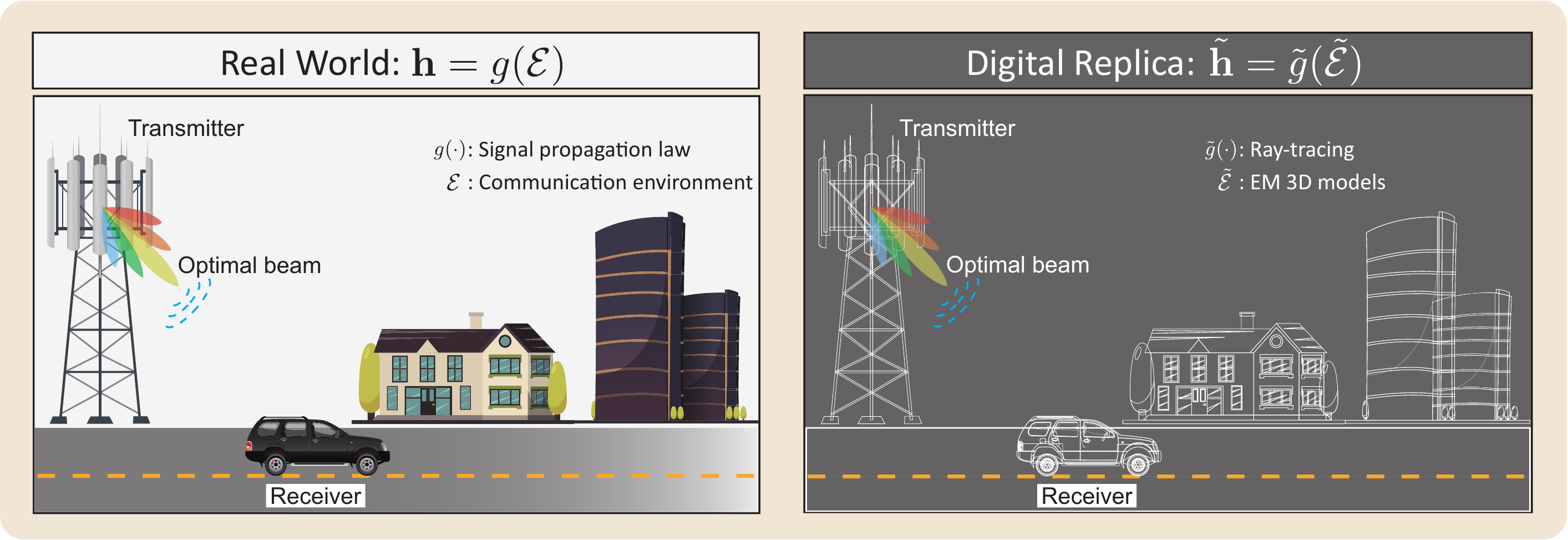}
		\caption{This figure shows a real-world communication system and its digital replica. In the real world, the channel between the transmitter and the receiver is determined by the communication environment and the signal propagation law. The digital replica employs the 3D model to approximate the communication environment and the ray tracing to model the signal propagation.}
		\label{fig:key_idea}
	\end{figure*}

	The digital replica relies on accurate ray tracing to simulate the channels. However, this ray tracing requires a large computational overhead. Machine learning (ML) models have demonstrated promising results in approximating complex functions with reduced computational overhead \cite{dong2019deep}. That motivates us to approximate the digital replica in a data-driven approach with ML, and solve the real-world communication tasks in an end-to-end manner. To further improve the ML performance, we propose to apply transfer learning and fine-tune the ML model with a small amount of real-world data.
	
	\iffalse
	Note that the concept of the digital twin has been investigated for network management and internet-of-thing systems \cite{khan2022digital}. However, this paper mainly focuses on exploiting the digital twin to aid physical layer operations of wireless communication systems.
	\fi
	Prior work has investigated the relevant sensing-aided wireless communication direction  \cite{charan2021vision, jiang2022lidar, Demirhan_mgazine_radar, ali2017millimeter}. In a sensing-aided communication setup, the communication devices (typically BSs) are equipped with various sensors. These sensors are used to capture sensing information about the surrounding environments, which are then utilized to aid assorted communication tasks \cite{charan2021vision, jiang2022lidar, Demirhan_mgazine_radar, ali2017millimeter}. For example, in \cite{charan2021vision, jiang2022lidar}, the authors employ camera and LiDARs at the BSs, and train ML models for the blockage detection and future beam prediction tasks. ML is particularly useful when the direct relationship between the sensing information (\textit{e.g.} camera and LiDAR) and the channel is hard to characterize. However, these ML-powered solutions may require large datasets that are hard to collect in reality.

	The proposed digital twin-based communications is a step forward from sensing-aided communications. The digital twins make full use of the sensing data by relating them to the communication channels through accurate ray tracing. The digital twin-based communications can potentially reduce or even eliminate the real-world channel acquisition overhead. Moreover, the digital twin can be straightforwardly extended to solve various communication tasks. The contribution of this paper can be summarized as follows: (i) We propose to approximate real-world communications using digital replicas based on accurate 3D model and efficient ray tracing. (ii) The digital replicas are exploited to simulate communication channels and solve real-world communication tasks with reduced/eliminated channel acquisition overhead. (iii) We approximate the digital replicas with ML models to further reduce the computational complexity. To evaluate the proposed digital twin aided wireless communication approaches, (v) we build a digital twin dataset comprising a real-world dataset (from DeepSense 6G \cite{DeepSense}) and a digital replica dataset based on accurate 3D ray-tracing.

	\section{System Model and Problem Formulation}
	In this paper, we consider a general MIMO communication system where $A$ base stations (BSs) serve $B$ mobile user equipments (UEs). The $a$-th $\left(a=\{1,\hdots,A\}\right)$ BS is equipped with an antenna array of $N_a$ elements. The $b$-th $\left(b=\{1,\hdots,B\}\right)$ UE employs an antenna array of $M_b$ elements. Moreover, we assume that the BSs have knowledge of the surrounding environments, which includes information about the positions, orientations, dynamics, shapes, materials of the BS, the UEs, and the other surrounding objects (that can act as reflectors/scatterers).
	\par	
	Without loss of generality, let $\bH_{b,a}$ denote the channel between the $a$-th BS and the $b$-th UE. The channels between all the BSs and UEs can be then represented by a set $\cH=\left\{\bH_{b,a} \mid b=\{1,\hdots,B\}, a=\{1,\hdots,A\}\right\}$. Accurate information about the channels $\cH$ is crucial for realizing the potential gains of the MIMO systems; many essential communications tasks for MIMO systems, such as precoding, beamforming/beam tracking, handover, resource allocation, and interference coordination, requires full or partial channel information. Let $\cT$ denote the solution space of one such communication task that requires the channel information $\cH$. Let $T^\star$ denote an optimal solution (where $T^\star \in \cT$). Further, let $\cS_{\cT}$ denote an existing method that optimally solves the communication task given the channel information $\cH$. Then, $T^\star$ can be written as
	\begin{align}\label{eq:optim_general}
		T^\star = \cS_{\cT}\left( \cH \right),
	\end{align}
	While the channel information $\cH$ is vital for MIMO systems, obtaining this channel information often requires a large acquisition overhead (beam sweeping, channel training/feedback, etc.) that degrades the overall system efficiency.
	%which degrades the spectral efficiency and system performance. Furthermore, this performance degradation can become even infeasible for massive MIMO systems since the training overhead scales up with the antennas.
	\par
	The objective of this paper is to solve the communication task in \eqref{eq:optim_general} while eliminating (or significantly reducing) the channel acquisition overhead. To that end, we propose a novel research direction that aims to solve \eqref{eq:optim_general} by approximating the real world with a digital replica.
	\section{From the Real World to the Digital Replica}\label{sec:real-to-synth}
	\iffalse
	the optimal beam $\bff^\star$ is a function of the channel vector $\bh$. When the perfect knowledge of the communication channel $\bh$ is available, the optimal beam can be solved using various methods, \textit{e.g.}, the exhaustive search.
	\fi
	In the real world, the communication channels $\cH$ are determined by the following two key components: The communication environment including the positions, orientations, dynamics, shapes, and materials of the BS, the UE, and other objects (reflectors/scatterers) in the surroundings, and (ii) the laws governing the wireless signal propagation phenomena.
	Let $\cE$ denote the communication environment and $g(\cdot)$ denote the signal propagation law, the communication channels can then be written as
	\begin{align}\label{eq:real_world}
		\cH = g(\cE).
	\end{align}
	When the communication environment $\cE$ and the signal propagation law $g(\cdot)$ are known, the solution $T^\star$ to the communication task can be obtained by substituting \eqref{eq:real_world} into \eqref{eq:optim_general}~as:
	\begin{align}\label{eq:optim_real}
		T^\star = \cS_{\cT}\Big( g(\cE) \Big).
	\end{align}
	Nevertheless, the precise ground-truth communication environment $\cE$ is difficult to obtain, and the exact expression of the signal propagation law $g(\cdot)$ remains unclear in complex environments. To that end, we propose to solve the communication task in \eqref{eq:optim_real} by approximating the communication environment $\cE$ and the signal propagation law $g(\cdot)$ in a digital replica. Particularly, we approximate the communication environment $\cE$ with Electro-magnetic (EM) 3D models and the signal propagation law $g(\cdot)$ with ray tracing.
	\par
	\textbf{EM 3D model} $\widetilde{\cE}$: The EM 3D models contain information about the positions, orientations, dynamics, shapes and materials of the BS, the UE, and other surrounding objects (reflectors/scatterers).
	Note that this 3D model information can be obtained using several approaches. (i) For static and fixed objects such as the neighboring buildings, the BS can memorize their position, shapes, and materials information since these objects are not likely to change frequently. (ii) In the context of sensing-aided information, the BSs can exploit various sensors such as cameras, radars, and LiDARs to obtain sensing information on both the surrounding stationery and dynamic objects. From this sensing information, the BSs can infer the position, orientation, dynamics, shapes, and materials of these objects. Moreover, once the BS identifies an object, the BS can use memorized and/or online data to refine the information about this object. For instance, if the BS detects a car of a certain model, the BS can search for the shape and material information of this car model in its or online database. (iii) Thanks to the recent development in the internet of things \cite{chettri2019comprehensive}, objects with communication capability can report/broadcast their information that are useful for the 3D model.
	\par
	\textbf{Ray tracing $\widetilde{g}(\cdot)$}: Based on the information in the 3D models, the channels $\cH$ can be modeled using stochastic and deterministic channel models. The stochastic channel models assume that the propagation parameters such as pathloss, delay spread, and angle spread, follow certain probability distributions. However, it is difficult to define these probability distributions for a specific scenario. By contrast, deterministic channel modeling methods like ray tracing do not rely on assumptions about the probability distributions of the propagation parameters. Instead, the ray tracing attempts to track the propagation paths between each transmit-receive antenna pair based on the geometry and material information in the 3D models, which preserves the spatial and temporal consistency. In this process, multiple propagation paths are explicitly modeled by considering various propagation effects, such as transmission, reflection, scattering, and diffraction. For each propagation path, the ray tracing produces path parameters including path gain, propagation delay, and propagation angles.

	These propagation path parameters generated by the 3D model and ray tracing can then be exploited to construct the channels in the digital replica. The channel impulse response $\widetilde{h}(t)$ between a transmit-receive antenna pair in the digital replica can be written as the sum of all the $L$ multi-path components, which is given by
	\begin{align}\label{eq:ray_tracing}
		\widetilde{h}(t) = \sum_{l=1}^{L}\alpha_l \delta(t-\tau_l) G_{\mathrm{t}}(\theta^{\mathrm{AoD}}_l) G_{\mathrm{r}}(\theta^{\mathrm{AoA}}_l),
	\end{align}
	where $\alpha_l$ and $\tau_l$ represent the complex gain and propagation delay of the $l$-th path. The angle of arrival and angle of departure are denoted by $\theta^{\mathrm{AoA}}_l$ and $\theta^{\mathrm{AoD}}_l$. $G_{\mathrm{t}}$ and $G_{\mathrm{r}}$ are the radiation patterns of the transmit and receive antennas.
	\par
	With accurate 3D model $\widetilde{\cE}$ and ray tracing $\widetilde{g}(\cdot)$, the solution $T^\star$ in \eqref{eq:optim_real} can be approximated using the solution obtained from the digital replica, $\widetilde{T}^\star$, as shown by
	\begin{align}\label{eq:optim_digital}
		\widetilde{T}^\star = \cS_{\cT}\Big( \widetilde{g}(\widetilde{\cE}) \Big).
	\end{align}
	To investigate the accuracy of the digital replica,	we define $s(T^\star, \widetilde{T}^\star)$ as the similarity function of $\widetilde{T}^\star$ and $T^\star$. The accuracy requirements on the 3D model and ray tracing can vary in different communication configurations and tasks. For instance, the channel state information prediction task in the sub-6GHz band may require more accurate 3D models and ray tracing than the beam prediction task in the mmWave band.

	One potential challenge of the digital twin lies in the high computational complexity of accurate ray tracing. This complexity can even increase when the 3D models have more details and contain a large number of interacting objects. As a result, the digital twin can suffer from high latency, which makes it less suitable for real-time applications. Next, we exploit ML to reduce such computational complexity.
	
	\section{Approximating the Digital Replica with Machine Learning}\label{sec:ML}
	The computational overhead and consequential latency of \eqref{eq:optim_digital} limits the feasibility of the digital replicas in real-time applications. Therefore, it is interesting to design a function $f(\cdot)$ that processes $\widetilde{\cE}$ and approximates the solution in \eqref{eq:optim_digital} with lower computational complexity and latency. In this paper, we take a data-driven approach and learn $f(\cdot)$ with ML.
	\par
	Let $f(\cdot\,;\Theta)$ denote an ML model with $\Theta$ representing the model parameters, the ML model is developed to learn a mapping function that takes in $\widetilde{\cE}$ and produces a solution $\widehat{T}$ that approximates the $\widetilde{T}^\star$ in \eqref{eq:optim_digital}. The objective of the ML optimization problem can be written as
	\begin{align}\label{eq:ml_target}
		\max_{f(\cdot\,;\Theta)} \bbE_{\widetilde{\cE}\sim\widetilde{\Upsilon}} \left\{ s(\widetilde{T}^\star, \widehat{T}) \mid \widetilde{\cE} \right\},
	\end{align}
	where $\widehat{T}$ is the output of the ML model $f(\cdot\,;\Theta)$ given the 3D model $\widetilde{\cE}$. $\widetilde{\Upsilon}$ denotes the underlying probability distribution of the 3D models. $\bbE\{\cdot\}$ is the expectation operator. The optimal ML model $f^\star(\cdot\,, \Theta^\star)$ that solves the ML optimization problem in \eqref{eq:ml_target} can be written as 
	
	\begin{align}\label{eq:optm_ml}
		f^\star\left(\cdot \, ; \Theta^\star\right) = \underset{f(\cdot \, ; \Theta)}{\argmax} \ \bbE_{\widetilde{\cE}\sim\widetilde{\Upsilon}} \left\{ s\left(\widetilde{T}^\star, f\left(p(\widetilde{\cE}); \Theta\right)\right)\right\},
	\end{align}
	where $p(\widetilde{\cE})$ is a function that extracts useful features from the 3D model; depending on the communication configurations and tasks, not all information in the 3D model $\widetilde{\cE}$ is useful. Therefore, we only input the useful features $p(\widetilde{\cE})$ to the ML model.
	
	The optimal ML model $f^\star(p(\widetilde{\cE}); \Theta^\star)$ can be obtained via a supervised learning approach. First, we randomly sample a total number of $D$ 3D models from the 3D model distribution $\widetilde{\Upsilon}$. For the $d$-th 3D model sample $\widetilde{\cE}_d$, we calculate the corresponding solution $\widetilde{T}^\star_d$ to the communication task $\cT$ using \eqref{eq:optim_digital}. This way, we can construct dataset of $D$ data points, and the $d$-th data point can be written as $\left(\widetilde{\cE}_d, \widetilde{T}^\star_d\right)$. Then, we train the ML to minimize a loss function on this dataset. The loss function measures how much the ML model approximation defers from the digital replica solution, which is given by 
	\begin{align}\label{eq:loss}
		J_{\mathrm{train}} =  -\sum_{d=1}^D s\left(\widetilde{T}^\star_d, f\left(p(\widetilde{\cE}_d); \Theta\right)\right),
	\end{align}
	Since the training data for the ML model $f\left(p(\widetilde{\cE}); \Theta\right)$ is generated from the digital replica, a large amount of training data can be relatively accessible for the training process. However, when the ML model is trained solely on the data generated by the digital replica, the ML model can be biased by the impairments in the 3D model and/or the ray tracing. In this case, the performance of the ML approximation is limited by the performance of the digital replica solution. To that end, a small amount of real-world data can be utilized to fine-tune the ML model that is trained on the digital replica dataset. This transfer-learning process is expected to calibrate the impairments brought by the digital replica and improve the performance of the ML model. The loss function for the transfer learning can be similarly written as \eqref{eq:loss}.
	
	\section{Case Study: Position-aided Beam Prediction}
	To verify the efficacy of the digital twin-based communications, we conduct an initial case study where the UE position is exploited for the mmWave beam prediction task.

	In our case study, we consider a communication system where a BS with $N$ antenna element is communicating with a single-antenna UE at the mmWave band. For simplicity, we assume the channel between the BS and the UE satisfies the narrowband block-fading channel model. In the downlink transmission, the BS sends a complex symbol $s$ using the beamforming vector $\bff \in \bbC^{N\times 1}$. The downlink receive signal at the UE can be written as
	\begin{align}\label{eq:signal_model}
		y = \bh^{H}\bff s + n,
	\end{align}
	where $\bh \in \bbC^{N\times 1}$ is the channel vector between the BS and the UE. 
	The transmitted symbol $s$ is constrained by the constant average transmit power $P$, \textit{i.e.}, $\bbE[|s|]^2 = P$. The $n$ represents the additive white Gaussian noise and satisfies $n \sim \cC \cN(0, \sigma^2)$. We assume that the BS adopt a beamforming codebook $\boldsymbol{\cF} = \{\bff_1, \hdots, \bff_Q\}$ incorporating $Q$ pre-defined beamforming vectors.
	To account for the constant-modulus constraint due to the analog beamforming architecture, the beamforming vectors in $\boldsymbol{\cF}$ satisfy $\|\bff_ q\|^2=1,\, \forall q\in\{1,\hdots ,Q\}$.
	\par
	Let $\cT_{pr}$ denote the beam prediction task solution space whose objective is to find the optimal beam $\bff^\star$ out of the codebook $\boldsymbol{\cF}$. The optimal solution of the beam prediction task $\cT_{pr}$ given the channel $\bh$ can be obtained by maximizing the receive SNR, which is given by
	\begin{align}\label{eq:optim_index}
		q^\star = \cS_{\cT_{pr}}(\bh) = \underset{q\in\{1,\hdots,Q\}} {\argmax}\,\left|\bff_q^H\bh\right|^2,
	\end{align}
	where $q^\star$ denotes optimal beam index. Substituting the 3D model $\widetilde{\cE}$ and the ray tracing $\widetilde{g}(\cdot)$ into \eqref{eq:optim_index}, the beam prediction solution obtained from the digital replica can be written as
	\begin{align}\label{eq:optim_index2}
		\widetilde{q}^\star = \underset{q\in\{1,\hdots,Q\}} {\argmax}\,\left|\bff_q^H\widetilde{g}(\widetilde{\cE})\right|^2.
	\end{align}
	Given the two beam indices $q^\star$ and $\widetilde{q}^\star$, their similarity function can then be defined using the Kronecker delta function $s(q^\star, \widetilde{q}^\star) = \delta({q^\star- \widetilde{q}^\star})$.
	Since we focus on the position-aided beam prediction task in our case study, the feature extraction function for the ML model is defined as $p(\cE) = \bu$, with $\bu$ denoting the position of the UE. 
	\iffalse
	\begin{table*}[]
		\centering
		\linespread{1.3}
		\small
		\caption{The Analogy and Notations in the Real World, the Digital Replica, and the ML Approximation.}
		\begin{tabular}{|l|l|l|}
			\hline
			Real World                          & Digital Replica                                & ML Approximation             \\ \hline
			Communication Environment $\cE$     & 3D model         $\widetilde{\cE}$                   & 3D model      $\widetilde{\cE}$                 \\ \hline
			Signal Propagation Law  $g(\cdot)$  & ray tracing     $\widetilde{g}(\cdot)$             & \multirow{3}{*}{\begin{tabular}[c]{@{}l@{}}ML Solution\\ $\widehat{T}=f\left(p(\widetilde{\cE}); \Theta\right)$\end{tabular}}\\ \cline{1-2}
			Communication Channel   $\bh=g(\cE)$    & Communication Channel $\widetilde{\bh}=\widetilde{g}(\widetilde{\cE})$ &                           \\ \cline{1-2}
			Communication Task Solution $T^\star = \cS_\cT\big(g(\cE)\big)$  & Communication Task Solution $\widetilde{T}^\star = \cS_\cT\big(\widetilde{g}(\widetilde{\cE})\big)$ &   \\ \hline
		\end{tabular}
	\end{table*}
	\fi
	\section{Experimental Setup}\label{Experimental Setup}
	In this section, we explain the real-world dataset, digital replica dataset, and the neural network (NN) adopted to evaluate the proposed digital-twin aided communications.
	\subsection{Real-World Data: DeepSense Scenario 1}\label{sec:real data}
\begin{figure}[t]
		\centering
		\includegraphics[width=0.9\linewidth]{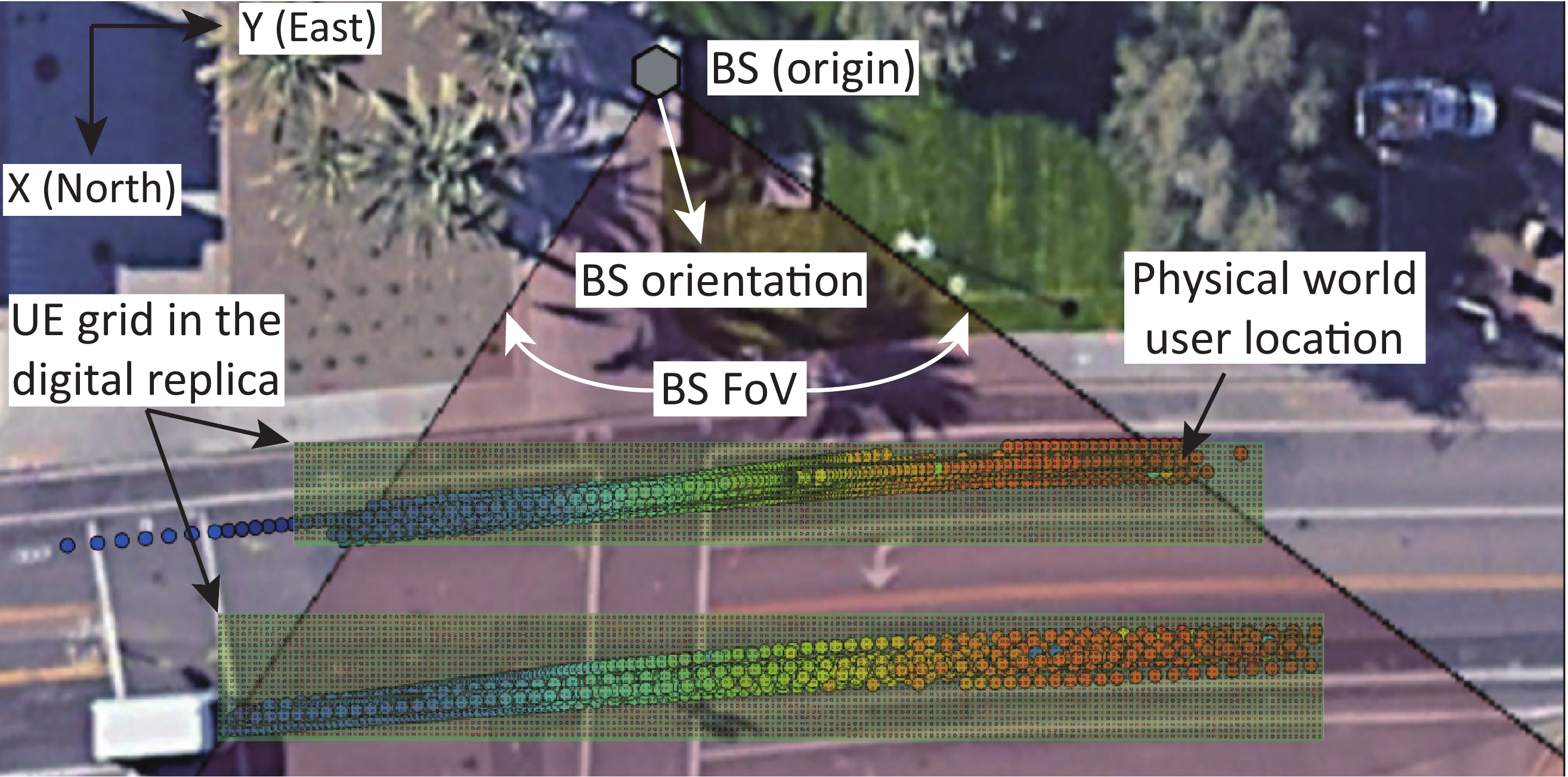}
		\caption{This figure shows the geometry layout of the real-world (Scenario 1 from DeepSense \cite{DeepSense}) data collection environment with the 3D model of the digital replica overlaid on the same layout.}
		\label{fig:scene}
	\end{figure}

	We adopt the DeepSense Scenario 1 for the real-world dataset \cite{DeepSense}. The data collection testbed incorporates a UE (vehicle) and a static BS. The UE carries a GPS sensor, and an omnidirectional  $60$ GHz mmWave transmitter. The BS uses a 16-element uniform linear array to receive the signals from the UE. The BS adopts a beam-steering codebook with $16$ beamforming vector. In the data collection process, the vehicle passes the two streets in front of the BS. At each time step, the BS performs a full beam sweeping to measure the receive powers of all 16 beams, and the UE captures its GPS position. The BS beam power and UE position measurements are synchronized. The $i$-th data point can be written as $(\bp^i, \bu^i)$, where $\bp^i \in \bbR^{16 \times 1}$ contains the power measurements of the $16$ beams and the $\bu^i \in \bbR^{2 \times 1}$ denotes the UE position. \figref{fig:scene} presents a visualization of all data points with the color indicating the beam of the highest power.
	
	\iffalse
	\begin{figure}[t]
		\centering	\includegraphics[width=0.9\linewidth]{setup.png}
		\caption{This figure shows the testbed for the real-world data collection. The BS receives the signal transmitted by the mobile UE (vehicle) on the 60 GHz mmWave band. The mobile UE carries a GPU sensor to record its position.}
		\label{fig:testbed}
	\end{figure}
	\fi
	
	\subsection{Digital Replica (Synthetic) Data}
	We build a digital replica of the real-world data collection environment. \figref{fig:scene} shows the 3D model of the digital replica overlaid onto the real-world environment. The digital replica consists of a BS and two UE grids. The UE grids cover the area where the UE can appear, \textit{i.e.}, the two streets in the real-world environment. The UE grids is discretized into multiple candidate UE positions with the spacing of $0.1$ meter. The channel between each candidate UE position and the BS is simulated using ray tracing. From the propagation path parameters produced by the ray tracing, the channel $\widetilde{\bh}^d$ between the $d$-th candidate UE position and the BS can be obtained by \eqref{eq:ray_tracing}. Then, for every candidate UE position, we can compute the power of all the beams in a pre-defined codebook. Lastly, for the $d$-th candidate UE position, we obtain a synthetic data point $(\widetilde{\bp}^d, \widetilde{\bu}^d)$ consisting of the beam power vector and the position. Note that, due to the 60 GHz frequency and the geometry, the channels in the real-world data are dominated by the line-of-sight (LoS) components. Therefore, in the digital replica, we also focus on the LoS paths and neglect the reflectors/scatterers. 
	\subsection{ML Model}
	We employ a fully connected NN architecture incorporating two hidden layers. Each hidden layer has 256 nodes and applies the ReLU activation. The input to the network is the position the UE in both Cartesian and polar coordinates. Although this input produces repeated information, we found that they lead to more stable ML performance with our dataset. To normalize the input data, we re-scale the distance in the polar coordinates by the maximum distance, and re-scale the $x$ and $y$ in the Cartesian coordinates by a shared maximum absolute value. The output layer adopts the standard classification setting, $\textit{i.e.}$, it has $16$ nodes and employs the softmax activation function. Each node produces a confidence score for one beam in the $16$-beam codebook being the optimal beam. We employ the Adam optimizer, and the learning rate is set to $1\times 10^{-2}$ and $1\times 10^{-4}$ for the training and fine-tuning. 
	\iffalse
	One motivation for adopting the ML model is to approximate the digital replica with reduced computational overhead and latency. Therefore, as shown by \eqref{eq:loss}, we train the above fully connected NN on the synthetic data points $(\widetilde{\bff}_d^\star, \widetilde{\bu}_d)$ generated by the digital replica. We use an initial learning rate of $1\ \times 10^{-2}$ and reduce the learning rate after every 20 epochs.
	\fi
	\iffalse
	\begin{table}[]
		\centering
		\caption{The Hyper-parameters of the NN Model}\label{tb:nn}
		\begin{tabular}{ll}
			\toprule
			\textbf{Hyper-parameters} & \textbf{Values}\\
			\midrule
			Input Size                & $4$ (Cartesian and polar coordinates)\\
			Output Size               & $16$ (beams)\\
			Layer Dimensions          & $\{4,256,256,16\}$\\
			Activation Function       & ReLU (hidden), Softmax (output)\\
			Optimizer                 & Adam\\
			Training Batch Size       & $32$\\
			Training Learning Rate    & $1\times 10^{-2}$\\
			Training Epochs           & $80$\\
			Fine-tuning Batch Size    & $8$\\
			Fine-Tuning Learning Rate & $1\times 10^{-4}$\\
			Fine-Tuning Epochs        & $40$\\
			Learning Rate Scheduling  & $0.2$ step decay at every 20 epochs\\
			\bottomrule
		\end{tabular}
	\end{table}
	\fi
	\iffalse
	\begin{figure}[t!]
		\centering
		\includegraphics[width=1\linewidth]{ray_tracing.pdf}
		\caption{This figure shows (in top-view) the 3D model of the BS and the UEs in the digital replica of \figref{fig:scene}.}
		\label{fig:ray_tracing}
	\end{figure}
	\fi
	\section{Evaluation Results}\label{Result}
	In this section, we evaluate the beam prediction performance of the proposed digital twin and ML approaches. We adopt two performance metrics: (i) The top-$k$ accuracy is defined as the percentage of the test data points whose ground-truth optimal beam lies in the predicted $k$ beams with the highest scores. (ii) The top-$k$ relative receive power measures the ratio between the highest receive power achieved by the top-$k$ predicted beams and the receive power of the ground-truth optimal beam.
	\begin{figure}[t!]
		\centering
		\includegraphics[width=0.9\linewidth]{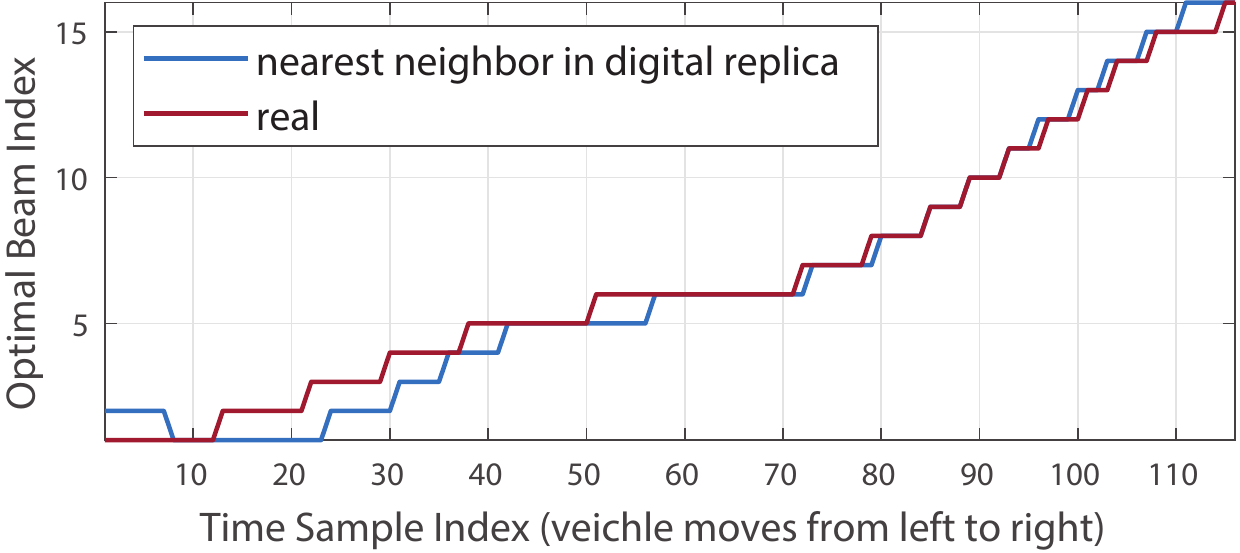}
		\caption{This figure shows the optimal beam indices of the real-world data points and their nearest neighbors as the UE passes by the BS.}
		\label{fig:optimal_beams}
	\end{figure}
	\begin{figure}[t!]
		\centering
		\includegraphics[width=0.9\linewidth]{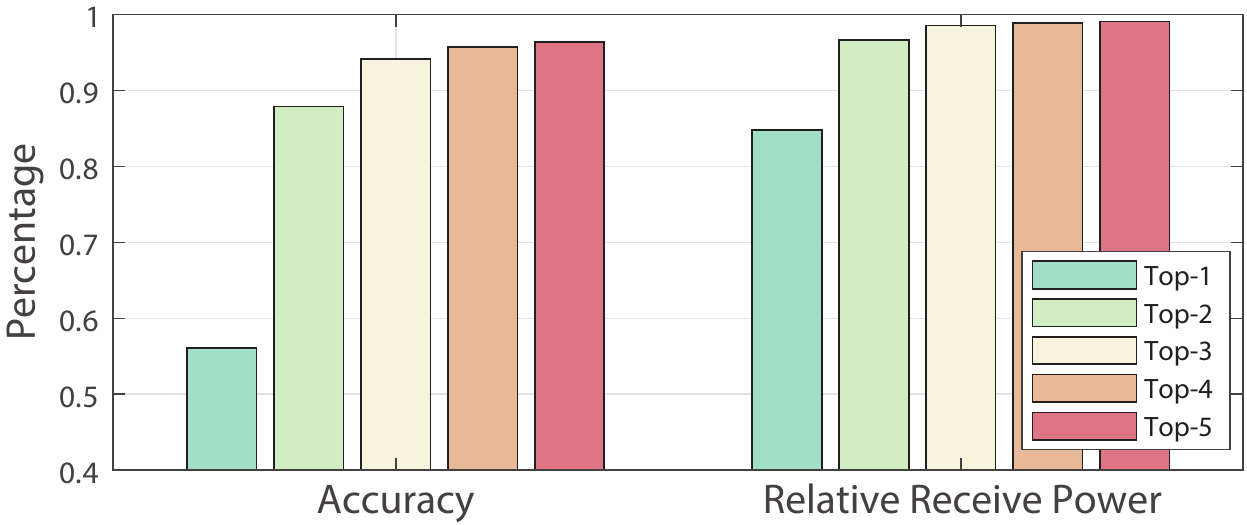}
		\caption{This figure shows the top-$k$ accuracy and relative receive power performance obtained by the nearest neighbor in the digital replica.}
		\label{fig:synth_acc}
	\end{figure}
	\subsection{Does the Digital Twin Matches the Reality?}
	The key idea presented in this paper is to utilize the digital replicas to simulate real-world communication systems. Here, we first investigate the accuracy of the digital twin in the context of the position-aided beam prediction task. Ideally, given a position, the optimal beam obtained in the real-world dataset should be similar to that indicated by the digital twin. As discussed in \sref{Experimental Setup}, each data point in the real-world or digital twin dataset contains a measured beam power vector and a UE position vector $(\bp, \bu)$. For each of the real-world data points $(\bp^i, \bu^i)$, we find its nearest neighbor $(\widetilde{\bp}^d, \widetilde{\bu}^d)$ in the digital replica, \textit{i.e.}, such that the distance between $\bu^i$ and $\widetilde{\bu}^d$ is minimized. Since the interval of the UE grid in the digital replica is small ($0.1$ m), the nearest neighbors in the digital replica can be considered an accurate approximation of the real-world data points. %For all the data points, we extract the optimal beam that provides the highest power.
	In \figref{fig:optimal_beams}, we compare the optimal beam indices of the real-world data points and their nearest neighbors in the digital twin as the UE passes by the BS from the left to the right. It can be seen that the real-world optimal beam index matches well with the optimal beam index of the nearest neighbor in the digital replica. The maximum difference between the optimal beam indices obtain from the real world and the digital replica is two, which translates to around $10^\circ$ difference in beam angle.
	
	Next, we directly apply the optimal beams of the nearest neighbors in the digital replica to the corresponding data points in the real world. \figref{fig:synth_acc} presents the top-$k$ accuracy and relative receive power obtained by this nearest neighbor approach. It can be seen that, the top-$1$ accuracy and relative receive power are $56.1\%$ and $84.8\%$, respectively. Despite the relatively low top-1 accuracy, the high receive power implies that the nearest neighbor often provides sub-optimal beams with near-optimal receive power. \textbf{It is worth highlighting that, with the nearest neighbor approach, $\bf 84.8\%$ receive power can be obtained using the digital twin without any beam training or channel estimation.}
	\subsection{Can the Model Trained in the Digital World Work in Reality?}
\begin{figure}[t!]
		\centering
		\includegraphics[width=0.9\linewidth]{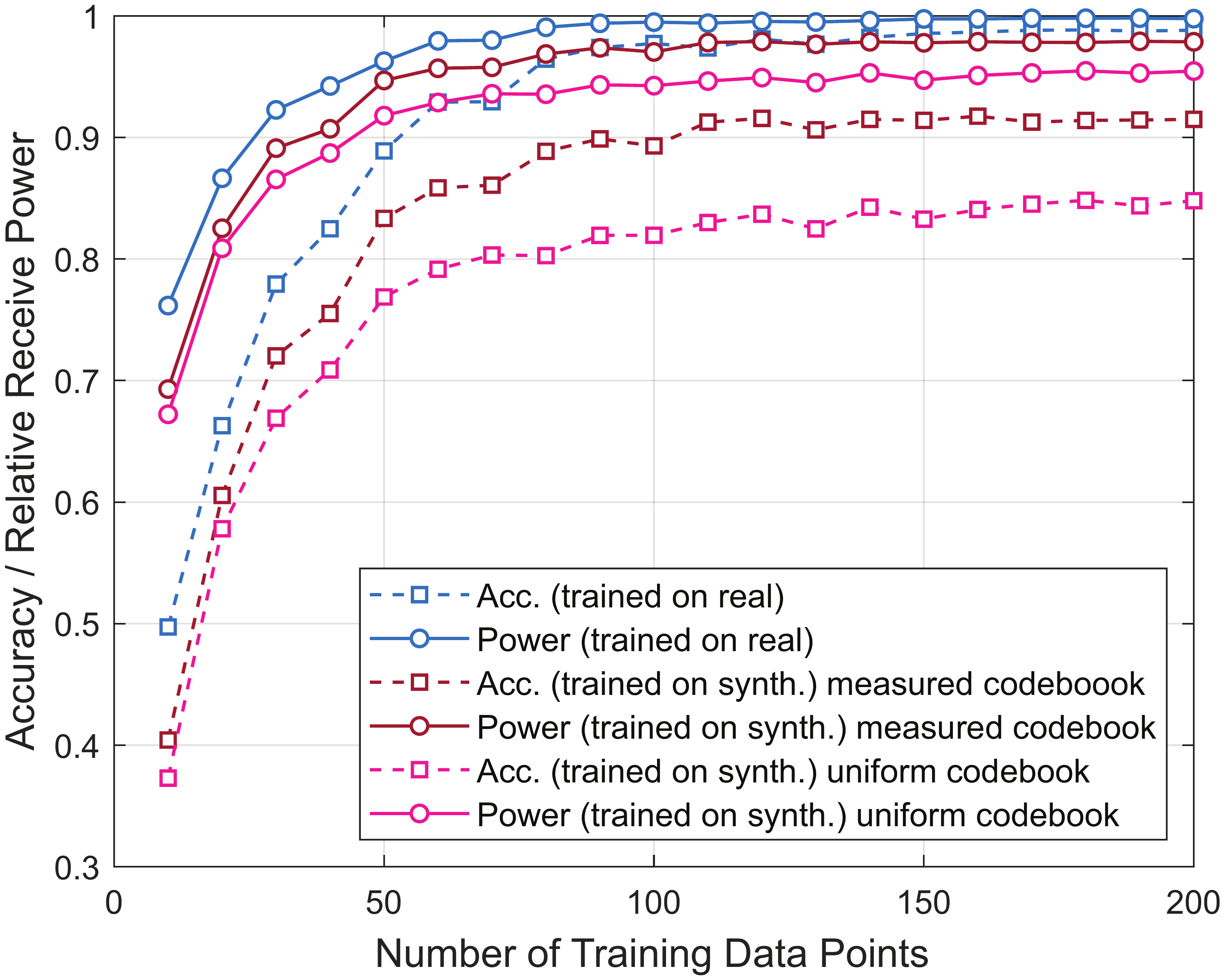}
		\caption{This figure shows the top-2 performance obtained by training the neural network on the real-world (real) or digital twin (synthetic) data and testing on the real-world data.}
		\label{fig:performance}
	\end{figure}

	\iffalse
	\begin{figure*}[t!]
		\centering
		\subfigure[The beam patterns of the BS array in DeepSense Scenario 1.]
		{
			\includegraphics[width=0.48\linewidth]{measured_codebook.pdf}\label{fig:a}
		}\hfill
		\subfigure[The beam patterns of the uniform beam steering codebook]
		{
			\includegraphics[width=0.48\linewidth]{uniform_codebook.pdf}\label{fig:b}
		}
		\caption{This figure shows the measured beam patterns of the BS array and the beam patterns of the uniform beam steering codebook.}\label{fig:codebooks}
	\end{figure*}
	\fi
	The large computational latency of the digital replica may not suit the need for real-time applications. Here, we investigate the performance of using the NN model to solve the digital replica simulation and the beam prediction task to reduce computation. In \figref{fig:performance}, we train the NN model on the real-world data or the digital replica data, and test the NN model on unseen real-world data. We experiment with two beamforming codebook in the digital replica, namely the measured codebook and the uniform beam codebook. In the measured codebook, we extract the beam angles of the beam patterns used in the DeepSense testbed, and construct a beam steering codebook with these angles. For the uniform beam codebook, we uniformly discretize the BS field-of-view into 16 angles and construct a DFT beam steering codebook accordingly. %\figref{fig:codebooks} presents the beam patterns of the BS array in DeepSense Scenario 1 and the beam patterns of the uniform beam steering codebook.

	It can be seen from \figref{fig:performance} that training on the digital replica data using the measured codebook can achieve a relatively good top-$2$ accuracy of $91.4\%$ when testing on unseen real-world data. \textbf{It is worth noting that this ML approach does not need any real-world training data.} Furthermore, the NN model can converge with only 100 data points from the digital replica, which demonstrates its high data efficiency. It can be also observed that the (inaccurate) uniform beam steering codebook decreases the top-2 accuracy to $84.8\%$. If we want to rely only on the digital replica data then we need very good modeling for important features such as the beamforming codebook.
	\subsection{Does Transfer-Learning Improve the ML Performance?}
\begin{figure}[t!]
		\centering
		\includegraphics[width=0.9\linewidth]{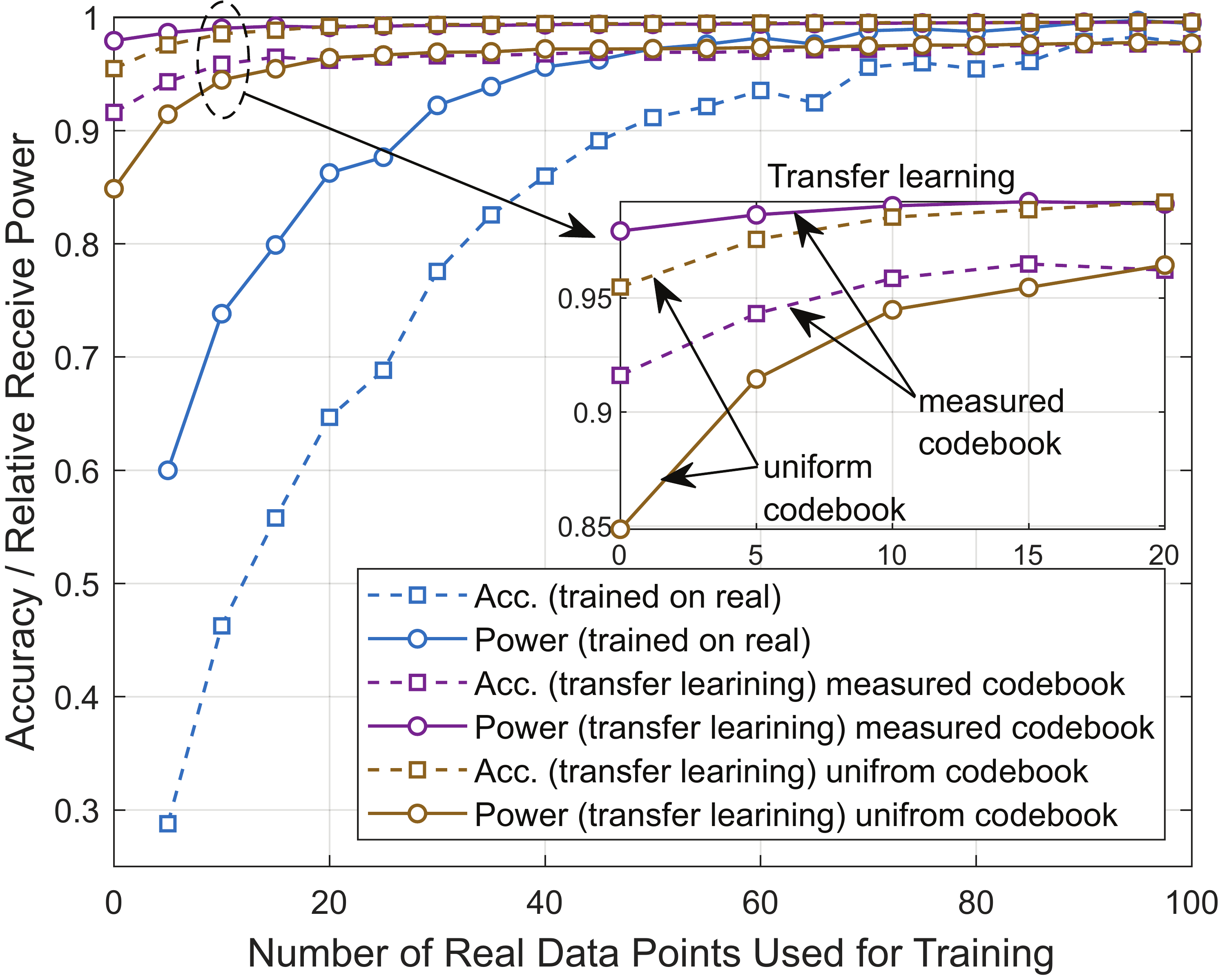}
		\caption{This figure shows the top-2 performance by applying transfer learning to the NN model using the measured and the uniform beamforming codebooks.}
		\label{fig:transer_performance}
	\end{figure}

	As discussed in \sref{sec:ML}, if the ML model is only trained on the data generated by the digital replica, the ML performance will be limited by the modeling impairments of the digital replica. To overcome this limitation, we propose to fine-tune the ML model on a small amount of real-world data after trained on the digital replica data. \figref{fig:transer_performance} shows the top-2 accuracy and relative receive power performance using this transfer learning approach. As a comparison, we show the performance achieved by directly training the NN on real-world data. It can be seen that a small number of real-world data points (less than $20$) can quickly improve the NN to achieve a near-optimal performance with the transfer learning gains. Directly training the NN on real-world data, however, requires around $100$ data points to converge to the optimal performance. This demonstrates that the digital twin can be used to pre-train ML models to efficiently save the cost and effort required by real-world data collection. The uniform beam steering codebook leads to lower performance compared to the measured beamforming codebook when fine-tuned on a very limited amount of real-world data. However, when the NN model is fine-tuned on more than $20$ real-world data points, the performance of the two codebooks become very similar. \textbf{This indicates that the mismatches between the real world and the digital replica can be efficiently calibrated by a small amount of real-world data.}
	\section{Conclusion}
	This paper proposes a novel direction that exploits the digital twin to aid wireless communications. Using the 3D model of the communication environment and ray tracing, the BS can simulate the communication system in a digital replica, which can be utilized to solve various communication tasks with reduced or eliminated channel acquisition overhead. Further, we propose to approximate the digital replica using ML with to reduce the computational overhead. Lastly, we propose to improve the ML model with a small amount of real-world data to compensate for the impairments in the digital replica. To evaluate the digital twin-based communications, we conduct an initial case study considering the position-aided beam prediction task.
	\iffalse
	In our experiment, the DeepSense Scenario 1 is adopted as the real-world collected dataset. We construct a digital replica to approximate the DeepSense Scenario 1 using accurate ray tracing. 
	\fi
	Experiments results can be summarized as follows. (i) The optimal beams in the digital replica can match the optimal beams in the real world. (ii) Training ML models on the digital replica data can achieve relatively good beam prediction performance. (iii) A small number of data points can quickly fine-tune the ML performance to near-optimal, and the digital twin can be used to save real-world training data. (iv) Mismatches between the digital replica and the real world can be calibrated by a small amount of real-world data.


\end{document}